# $TiO_2$ nanotubes: N-ion implantation at low-dose provides noble-metal-free photocatalytic $H_2$-evolution activity

Xuemei Zhou, [a] Volker Häublein, [b] Ning Liu, [a] Nhat Truong Nguyen, [a] Eva M. Zolnhofer, [c] Hiroaki Tsuchiya, [d] Manuela S. Killian, [a] Karsten Meyer, [c] Lothar Frey, [b] Patrik Schmuki *[a]

[a]Department of Materials Science WW-4, LKO, University of Erlangen-Nuremberg, Martensstrasse 7, 91058 Erlangen, Germany. Tel.: +49 91318517575, fax: +49 9131 852 7582. E-mail: schmuki@ww.uni-erlangen.de.

[b]Fraunhofer Institute for Integrated Systems and Device Technology IISB, Schottkystrasse 10, 91058 Erlangen, Germany.

[c]Department of Chemistry and Pharmacy, Inorganic Chemistry, Friedrich-Alexander University Erlangen – Nürnberg (FAU), Egerlandstr. 1, 91058 Erlangen, Germany.

[d]Division of Materials and Manufacturing Science, Graduate School of Engineering, Osaka University, 2-1 Yamada-oka, Suita, Osaka 565-0871 Japan.

**Abstract:** Low-dose nitrogen implantation induces in $TiO_2$ nanotubes a co-catalytic activity for photocatalytic $H_2$-evolution. The use of an ion implantation process leads to a N-implanted zone only at the top part of the tubes. The coupling of this top layer and the underlying non-implanted part of the nanotubes strongly contributes to an efficient carrier separation and thus to a significantly enhanced $H_2$ generation.

Ever since the groundbreaking work of Fujishima and Honda [1] in 1972, photocatalytic water splitting has attracted tremendous research interest. The process is the application of a simple photoelectrochemical principle: If light (preferably solar light) is irradiated on a semiconductor, electron-hole pairs will be generated, separated on conduction and valence band and may react at the semiconductor surface with red-ox couples in the surrounding (if the energetic positions of conduction and valence band relative to the corresponding red-ox potentials are suitable).[2] In spite of hundreds of investigations on a wide range of photocatalysts, $TiO_2$ still remains the most investigated semiconductive material for photocatalytic hydrogen generation from various electrolytes (with or without sacrificial agents), as it has a conduction band-edge level compatible with the generation of $H_2$ from water, is economically and ecologically sound, and especially as it has a very high (photo) corrosion resistance. [2, 3]

Nevertheless, a main setback of using titania is that the charge transfer reaction kinetics on plain $TiO_2$ surfaces and thus the $H_2$ production rates are extremely slow. [4] If not, co-catalysts are being used that typically are noble metals such as Pt, Pd or Au. [4, 5] These co-catalysts facilitate or stimulate the transfer of photogenerated electrons from the conduction band to the surrounding and thus promote hydrogen production. [6] The use of such a noble metal co-catalyst, however, puts into question the economic benefit of using cost-friendly $TiO_2$ as a base material.

Therefore pathways to create noble-metal-free $H_2$ evolution activity on $TiO_2$ are not only of considerable scientific but also of high economic interest. Up to now only comparably little efforts targeted treatments that can create an intrinsically increased reactivity of titania towards $H_2$ generation. [7] Nevertheless, for a number of photocatalytic reactions various defects in $TiO_2$ surfaces are widely recognized to be strongly affecting photocatalytic rates but also known to be unstable if not present in a subsurface configuration. [8]

A most efficient way to introduce a sub-surface configuration of a broad range of lattice defects (namely vacancy/interstitial pairs) into any crystalline material is high energy ion implantation. [13]

In the present work we introduce the use of low levels of nitrogen ion implantation into $TiO_2$ nanotube layers to modify them with a defined ion and defect distribution. These nanotubes (NTs) then are investigated for their photocatalytic $H_2$ generation under solar light illumination. The results show that N–implantation (and the accompanied damage), at sufficiently low doses, can effectively induce an activation of anatase $TiO_2$ NTs for noble-metal-free photocatalytic $H_2$ evolution (Figure 1).
We selected for implantation of nitrogen ions, not only as they allow the creation of highly controllable lattice damage but also as they were expected to potentially provide additional beneficial effects towards the electronic properties of the $TiO_2$ NTs. Nitrogen treatments of titania have mainly been investigated to modify (narrow) the band-gap of $TiO_2$. [14] Moreover, recent theoretical and experimental work [15] reported nitrogen (in substitutional and interstitial positions) to have a stabilizing effect on defect species such as ($Ti^{3+}/O_v$) by a charge transfer resonance with the nitrogen states.

For our experiments we used self-organized $TiO_2$ nanotube layers grown from titanium metal sheets by electrochemical anodization in an $NH_4F/H_2O$/EG electrolyte. Such layers are shown in Figure 1a and consist of individual tubes of 80 nm in diameter and a length of 6 μm. Alternatively some tube layers were grown to a length of ≈ 500 nm (more details are given in the supporting information). The NT-layers were then converted by air annealing at 450 °C to an anatase structure (as anatase is the most efficient polymorph of $TiO_2$ for photocatalytic $H_2$ generation [7b, 8a]).

Ion implantation was carried out with a Varian 350 D ion implanter at 60 keV at doses of $8\times10^{14}$ ions/cm$^2$ and $1\times10^{16}$ ions/cm$^2$, respectively. Figure 1b shows calculations of the implantation effects in the $TiO_2$ nanotube substrate in terms of depth profiles for nitrogen ions, as well as damage (by oxygen- and titanium- recoil) using TRIM code [13]. The calculations show that at 60 keV implantation a maximum for the nitrogen ion implant located at approximately 200 - 600 nm below the surface and a maximum point defect damage at ≈ 100 - 500 nm below the surface.

Note that the TRIM results indicate an important difference for the two lengths of tubes used in this work: For the shorter tubes (500 nm), the implant/damage profile penetrates the full tube length, whereas for the longer tubes (6 μm) ion implantation leads to a modification only in the top part of the tubes – i.e. in this case the implanted zone at the top and the underlying unaffected $TiO_2$ NTs may potentially establish an electronic homo-junction within the tube wall (illustrated in Figure 1b and discussed later) (see also details in the supporting information). Fig. 1c shows that ion implantation can have a strong effect on the photocatalytic $H_2$ production rate. Interestingly the tube layers that were exposed to a low-N-dose show a strongly enhanced activity, while for the tubes implanted with a higher dose virtually no or only a small effect can be seen. Moreover, the beneficial effect of low dose nitrogen implantation is much more pronounced for the longer $TiO_2$-NTs than for the short tubes (although both contain the same ion and damage doses and distributions). This strongly suggests photogenerated electrons in the intact part of the long tubes can reach the implanted zone and react there to form $H_2$ (we will discuss this point in more details later).

In order to characterize effects of the N-implantation on the structure of the NTs we studied the layers before and after nitrogen implantation by XRD, Raman and TEM. From XRD taken for tube layer as in Figure 1a, a clear decrease of the anatase main peak located at 25.2 ° can be seen after high dose N implantation (Figure 2a). This is in line with a partial amorphization of the tubes after implantation. [16] For the low dose nitrogen implantation, no apparent decrease of the anatase peak or broadening is evident from XRD spectra. Also evaluating the peak using the Scherrer equation we obtained a similar average characteristic crystallite length of 38.0 nm for reference nanotubes and 38.2 nm for low-dose implanted NTs, while for the higher-dose implanted tubes clear peak broadening is observed (29.0 nm). From Raman spectra (Figure 2b) for plain $TiO_2$ NTs a typical anatase Raman signature can be obtained. For the high dose implanted $TiO_2$ NTs a significant decrease of all modes is observed that can be ascribed to a partial amorphization of the anatase crystal structure [17] - this is well in line with XRD. For the low dose implant NTs, similar signatures with plain $TiO_2$ NTs can be observed, however, the peak intensity of all modes (Figure 2b inset) decreases slightly without a clear change in the position of the Raman band. Such an effect in Raman spectra has been reported for the introduction of defects in $TiO_2$ (such as the presence of an increased vacancy concentration [5b, 18]).

Based on the TRIM calculations (Figure 1b and Figure S1), one would expect the damage and implant concentration to peak at 200 - 600 nm below the surface. Therefore we studied the tubes at various depths with HRTEM and corresponding SAED. Figure 2c shows SAED patterns taken at different distances from the tube top for a 6 µm long nanotube. Even if the tube is implanted at a higher dose ($1\times10^{16}$ ions/cm$^2$), in the upmost zone the SAED image reveals a clear anatase signature. However, for SAED taken in the range of 400 nm – 600 nm below the top, the patterns reveal almost total amorphization of the tube walls. This amorphization in the maximum damage zone of high dose nitrogen implantation is also visible from HRTEM images provided in the SI (Figure S4). In contrast, for samples implanted at a lower dose even in the maximum damage zone, no changes in the reflex-patterns or blurring in the SAED is apparent – i.e. still a clear anatase pattern is obtained.

This indicates that under these conditions the introduced damage is not sufficient to significantly alter the basic anatase character of the tube wall. SAED for tube regions at 1.5 µm below the top and underneath shows again clear patterns of crystalline anatase for both implant doses.

Chemical composition analysis with XPS and TOF-SIMS confirms the presence of N in the implanted tubes (see supporting information, Figure S5 and discussion).

To characterize the tubes in view of a modification of the electronic and electrochemical properties we carried out impedance and photoelectrochemical measurements as shown in Fig. 3 and discussed in detailed in the SI.

Overall, the results of Figure 3a-d show that high dose nitrogen implantation has a strong detrimental effect on photocurrent harvesting and the charge transfer from the tubes to the surrounding electrolyte. These effects, to a large extent, can be ascribed to amorphization of the anatase structure by ion beam damage. Low dose N-implantation, on the other hand, leads to sub-band-gap states (≈ 0.2 eV below the conduction band) that drastically reduce the charge transfer resistance.

For longer tubes low dose ion implantation leads to a zone close to the tube top that shows these beneficial effects. In this case, photogenerated electrons from underlying intact tube parts can easily reach the active zone. (Considering that light in the band-gap region of anatase has an absorption depth into $TiO_2$ nanotubes of a few micrometers and an electron diffusion length in the range of several 10 micrometers have been reported for anatase $TiO_2$ NTs. [3b, 24])

Thus an electron harvesting/charge-transfer-activity combination is established that significantly contributes to the overall $H_2$ evolution efficiency. Electron transfer towards the implanted zone may further be facilitated by the fact that a beneficial electronic junction is formed. I.e., in the implanted zone the Fermi level lies closer to the conduction band (reflected by the different flat band potential (Figure 3a)). This leads at the meeting point of implant and implant-free location to a homo-junction in the tube wall (as illustrated in supporting information Figure S7) – this effect additionally drives photogenerated electrons towards the charge-transfer-active zone.

In order to gain additional information on the nature of the defects induced by low dose ion implantation we used continuous wave X-band electron paramagnetic resonance spectroscopy (EPR). EPR is sufficiently sensitive to characterize in $TiO_2$ very low concentrations of paramagnetic centers, such as $Ti^{3+}$ or oxygen vacancy ($O_v$) states [25]. Figure 4a shows CW X-band EPR spectra for the reference (upper spectrum) and the low dose nitrogen implanted $TiO_2$ nanotubes (lower spectrum) measured at 7K (liquid He). For plain $TiO_2$ NTs, only trace amounts of defects ("F-centers" or oxygen vacancies [26]) are detectable. For the implanted sample, an additional and significantly stronger signal is apparent, indicating the presence of a second type of paramagnetic centers. [15b] Simulation of the spectra allows for the characterization of two species, as shown in Fig. 4b: for both samples, one slightly rhombic signal with g-values of $g_1 = 2.0022$, $g_2 = 2.0015$, and $g_3 = 1.9461$ is present (subspecies A); and, only for the low dose nitrogen-implanted $TiO_2$ nanotubes, an additional second axial signal with g-values of $g_{||} = 2.0913$ and $g_{\perp} = 1.9120$ is apparent (subspecies B). This signature can be attributed to a $Ti^{3+}/N_b$• ($N_b$: bulk nitrogen in substitutional or interstitial position) combination in $TiO_2$ in accordance with literature and experimental work. This configuration stabilizes $Ti^{3+}$ centers by charge transfer resonance[15] and thus, such $Ti^{3+}$ species are regarded as particularly robust. Overall it is plausible to ascribe the beneficial effect (for $H_2$ evolution) observed by low dose N-implantation mainly to the formation of $Ti^{3+}$ defect states that are energetically situated close to the conduction band and are embedded in a crystalline anatase matrix. These defects may additionally be stabilized by the presence of nitrogen.

To summarize, low dose (sufficiently high energy) nitrogen implantation into anatase $TiO_2$ nanotubes leads to an intrinsic co-catalytic effect that strongly promotes the photocatalytic $H_2$ evolution performance of $TiO_2$ nanotube layers without the use of any (noble metal) co-catalyst. This is in contrast to high-dose nitrogen implantation that leads to amorphization of the implanted region. Using a sufficiently low dose allows to essentially maintain an intact anatase matrix dosed with distinct defects and dopants. This damage configuration is a key for the catalytic activity for $H_2$ evolution. The catalytic states are situated close to the conduction band and are apparent in capacitance measurements, photocurrent spectra, photocurrent transients and EPR spectra. A particular advantage of creating these defect centers by N-implantation is that the active centers can be placed as a coherent layer at a subsurface location determined by the selected energy – i.e. in vertically aligned $TiO_2$ nanotubes an active zone (junction) can be embedded in intact nanotubes and a highly synergistic combination of light/electron harvesting and co-catalytic activity can be established.

**Acknowledgements**

The authors would like to thank ERC, DFG and the EAM cluster of excellence for financial support. The authors would also like to thank Yuyun Yang for the Raman measurements.

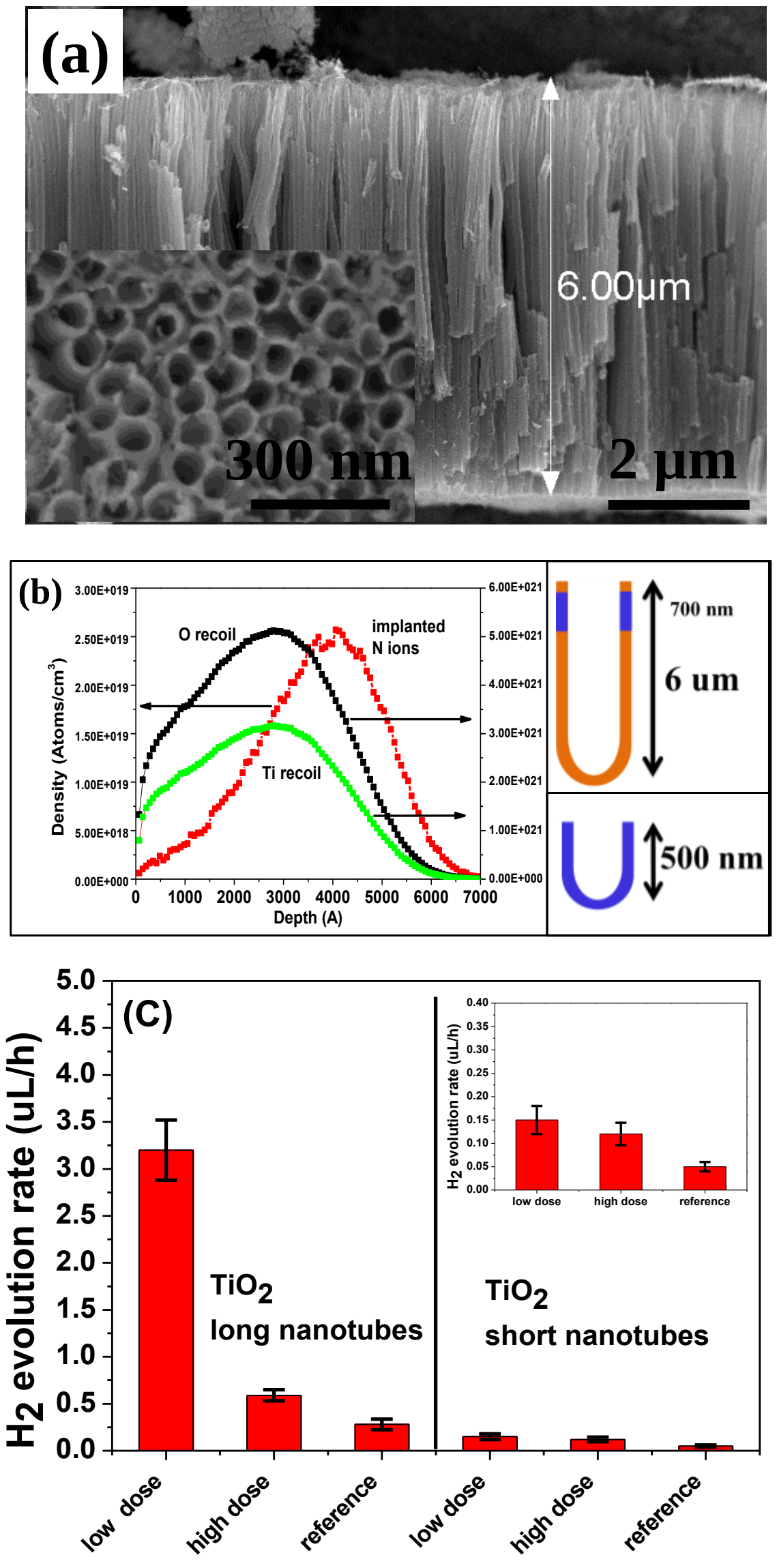

**Figure 1.** (a) SEM images of $TiO_2$ nanotube layers (~ 6 µm long) used in a cross section and as top view (insert). (b) TRIM simulation of depth-distribution of N ions and damage (Ti- and O recoil) into a $TiO_2$ nanotube target with an energy of 60 keV at a dose of $8\times10^{14}$ atoms/cm$^2$. (c) Photocatalytic $H_2$ evolution from $TiO_2$ – NTs long (6 µm, Fig.1a) and short (500 nm, Fig. S3) under AM 1.5 (100 mW/cm$^2$). Nitrogen implant is compared for two doses of $8\times10^{14}$ ions/cm$^2$ and $1\times10^{16}$ ions/cm$^2$ and bare (non-implanted) $TiO_2$ nanotube layer.

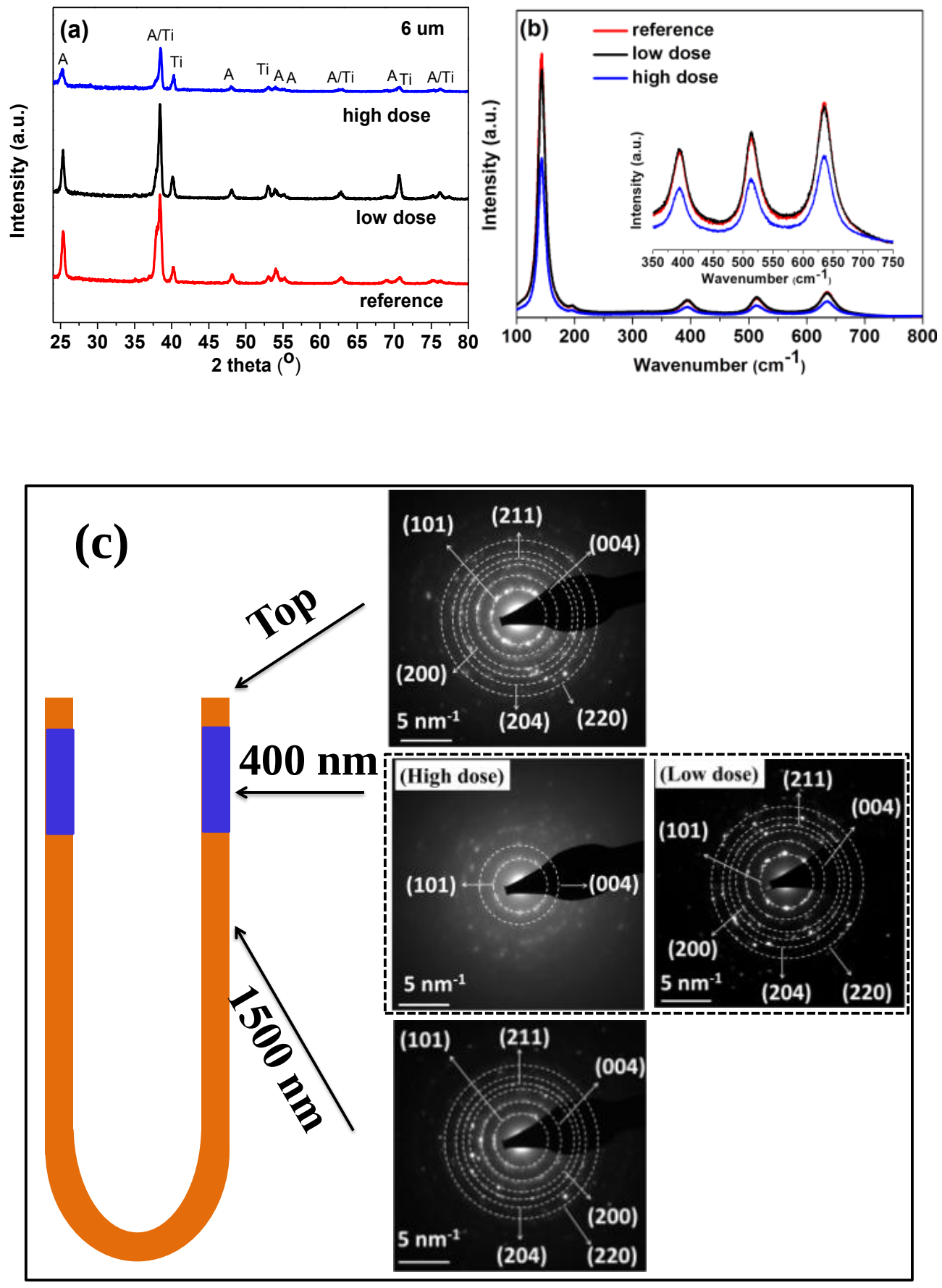

**Figure 2.** (a) X-ray diffraction (XRD) and (b) Raman spectra of $TiO_2$ nanotube layers (as in Fig. 1a) before and after N implantation with doses of $8\times10^{14}$ ions/cm$^2$ and $1\times10^{16}$ ions/cm$^2$, (c) SAED patterns of N implanted $TiO_2$ NTs with a dose of $1\times10^{16}$ ions/cm$^2$, taken at the top, 400 nm - 600 nm below the surface and 1500 nm below the surface showing amorphization in the maximum implant zone (100 - 500 nm below the surface): left image. And SAED at same level (400 nm – 600 nm) below surface for low dose ($8\times10^{14}$ ions/cm$^2$) implantation: right image – showing still an intact anatase pattern.

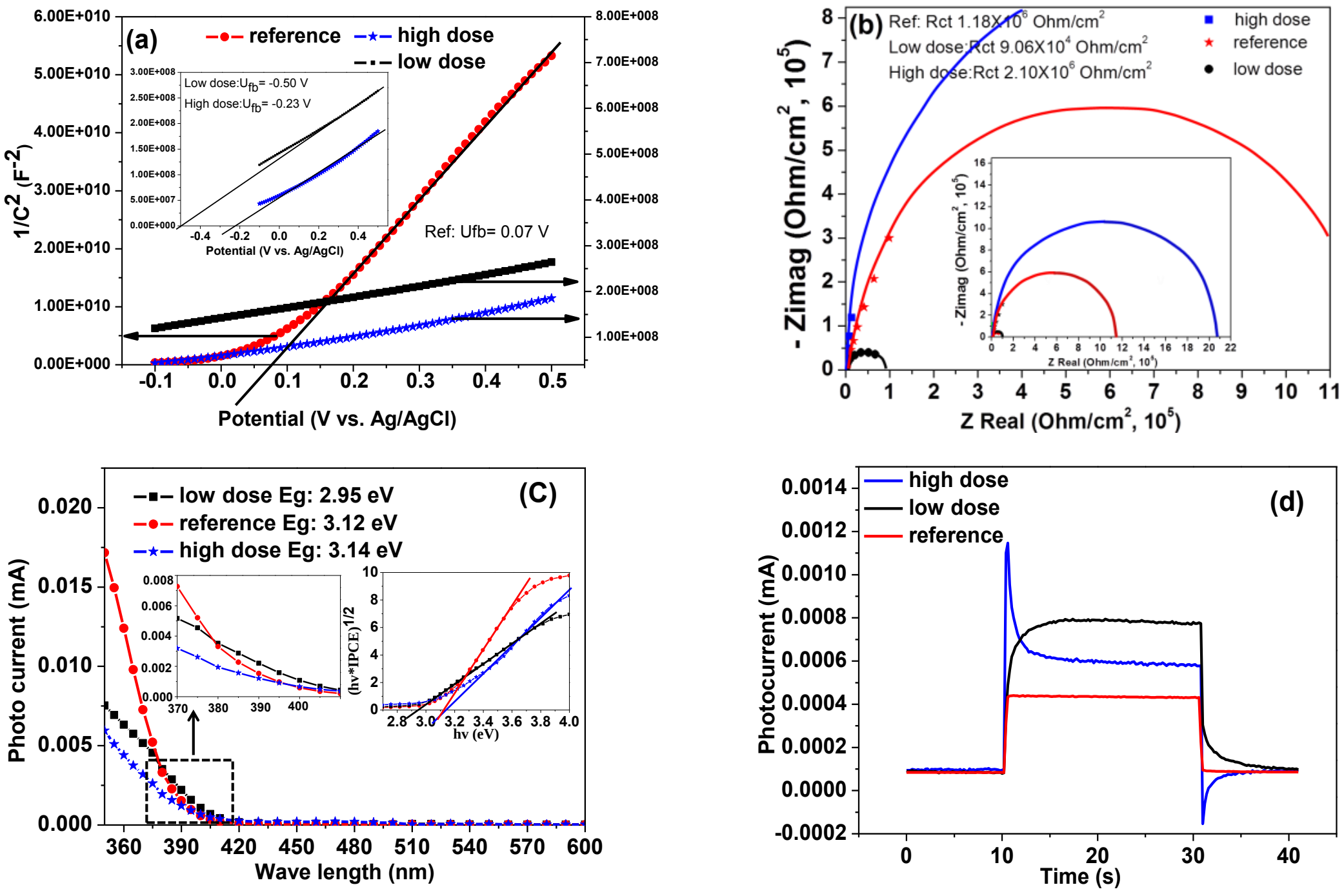

**Figure 3.** Electrochemical and photoelectrochemical characterization for the N-implanted and reference $TiO_2$ nanotubes. (a) Mott-Schottky plots (inset: enlarged plots for the low and high dose N implanted samples) obtained in 0.1 M $Na_2SO_4$. (b) Nyquist plots from the EIS measurements in 0.1 M $Na_2SO_4$ at 200 mV (vs. Ag/AgCl). The inset shows the full range Nyquist plot. (c) Photocurrent spectra in 0.1 M $Na_2SO_4$ at 500 mV (vs. Ag/AgCl) (inset: band gap valuation from the photocurrent spectra). (d) Photocurrent transients taken at 405 nm in 0.1 M $Na_2SO_4$ at 500 mV (vs. Ag/AgCl). All tubes were as in Figure S3.

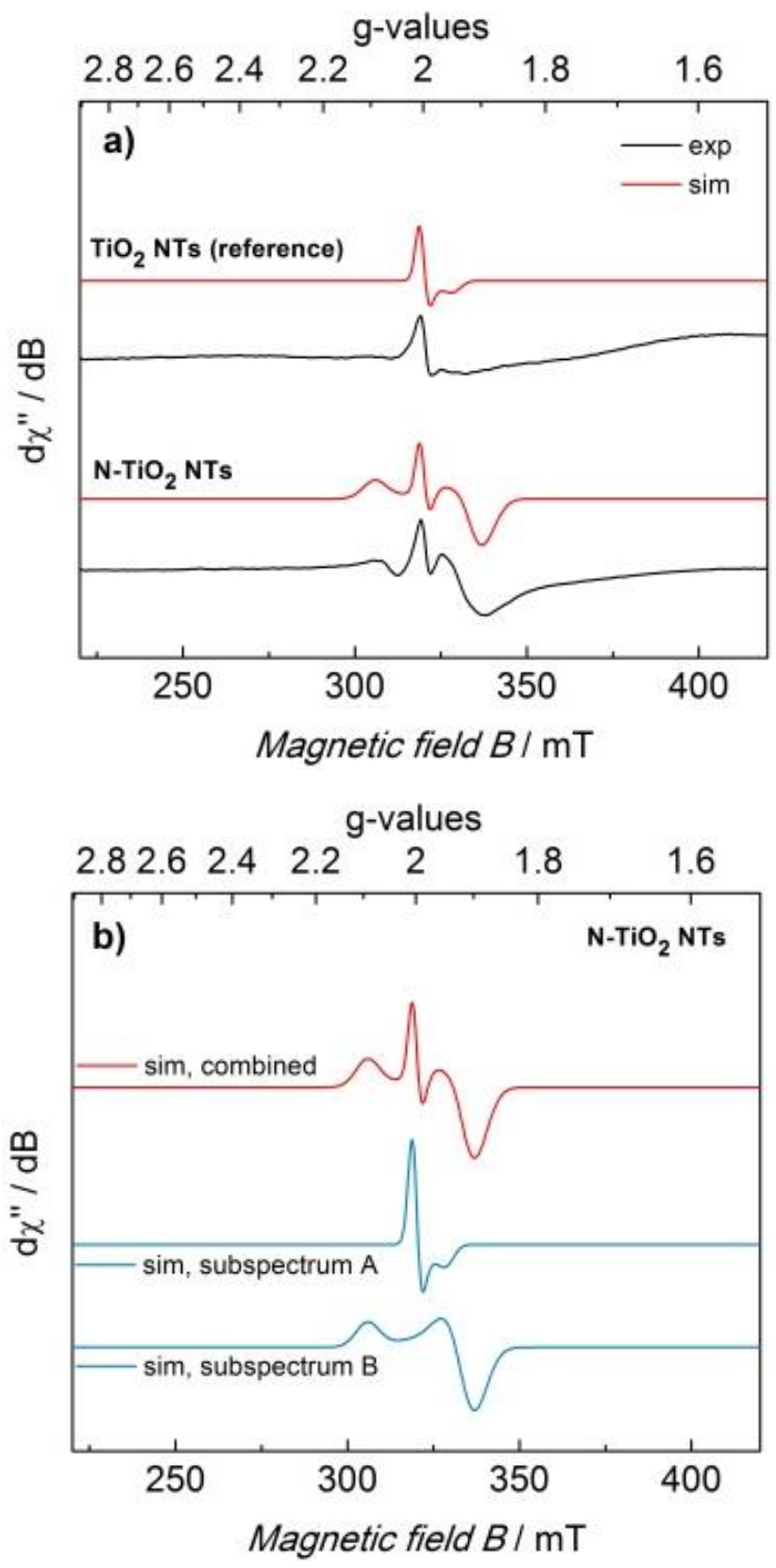

**Figure 4:** a) CW X-band EPR spectra of low-dose nitrogen-implanted $TiO_2$ nanotubes (N-$TiO_2$ NTs, lower spectrum) and, for comparison, an non-implanted reference (plain $TiO_2$ NTs, upper spectrum).

# Supporting Information

# $TiO_2$ nanotubes: N-ion implantation at low-dose provides noble-metal-free photocatalytic $H_2$-evolution activity

Xuemei Zhou[1], Volker Häublein[3], Ning Liu[1], Nhat Truong Nguyen[1], Eva M. Zolnhofer[2], Hiroaki Tsuchiya[4], Manuela S. Killian[1], Karsten Meyer[2], Lothar Frey[3], Patrik Schmuki[1]*

[1]Department of Materials Science WW-4, LKO, University of Erlangen-Nuremberg, Martensstrasse 7, 91058 Erlangen, Germany;

[2]Department of Chemistry and Pharmacy, Inorganic Chemistry, Friedrich-Alexander University Erlangen – Nürnberg (FAU), Egerlandstr. 1, 91058 Erlangen, Germany.

[3]Fraunhofer Institute for Integrated Systems and Device Technology IISB, Schottkystrasse 10, 91058 Erlangen, Germany;

[4]Division of Materials and Manufacturing Science, Graduate School of Engineering, Osaka University, 2-1 Yamada-oka, Suita, Osaka 565-0871 Japan.

*Corresponding author. Tel.: +49 91318517575, fax: +49 9131 852 7582*

*Email:* schmuki@ww.uni-erlangen.de

**Experimental section:**

**$TiO_2$ nanotubes (NTs) preparation:**

$TiO_2$ nanotubes (NTs) were grown by self-organizing anodization of a Ti sheet (Advent, 0.125 mm thickness, purity 99.6 +%). Before anodization the titanium foils were degreased by sonication in acetone and ethanol, respectively, rinsed with deionized water, and dried in a nitrogen stream. Tubes of a length of 6 – 7 µm were prepared by anodization in an electrolyte consisting of ethylene glycol (EG, Sigma–Aldrich), DI water (1M) and $NH_4F$ (0.2 M, Sigma–Aldrich, 98%) for 20 min at 60 V using a power supply (VOLTRAFT VLP2403Pro) in a two electrode configuration with a platinum counter electrode. Short $TiO_2$ nanotubes (≈ 500 nm layer thickness) were made by anodization in a water based electrolyte (1M $H_2SO_4$ (ACS reagent, 95.0-98.0%, Sigma-Aldrich) with 0.125 wt % HF (≥ 40%, Sigma-Aldrich)) using the same configuration as above. The potential was first ramped with a sweep rate of 50 mV/s to 20 V, then it was held for 5 h at 20 V. More details are given in [S1]. In order to convert the amorphous tube layer to anatase, thermal annealing was performed in air using a rapid thermal annealer (Jipelec JetFirst 100) at 450 °C for 3h with a temperature ramp rate of 0.5 °C/s. The as prepared $TiO_2$ nanotube layers were then used as targets for N ion implantation.

**Nitrogen Implantation**:

Nitrogen ($N^+$) ion implantation into the annealed $TiO_2$ NT layers was carried out using a Varian 350 D ion implanter at 60 keV acceleration energy. The nominal dose used was $8\times10^{14}$ ions/cm$^2$ (denoted as low dose) and $1\times10^{16}$ ions/cm$^2$ (denoted as high dose), respectively. Depth profile simulation was performed using TRIM-2013, $TiO_2$ (Target No. 652) was selected as target which was corrected for the average density of $TiO_2$ NTs of 1.19 g/cm$^3$.

**Characterization:**

A Hitachi FE-SEM S4800 was used for morphological characterization of the samples. XRD patterns were collected using an X'pert Philips PMD diffractometer with a Panalytical X'celerator detector, using graphite-monochromatized CuKa radiation ($\lambda$ = 1.54056Å). XRD peaks were evaluated using the Scherrer equation:

$$\mathrm{r} = \frac{K\lambda}{\beta \cos\theta}$$

where r is the average size of crystalline particles, K is the shape factor (here set as 1), λ is the wavelength of Cu $K_\alpha$ =1.54056Å, β is the line broadening taken as full width at half maximum (FWHM), θ is the Bragg angle (here the main peak at 25.2° is used, θ is thus 12.6°).

High-resolution transmission electron microscopy (HRTEM) and selected area electron diffraction (SAED) patterns were acquired with a Philips CM 300 UT TEM at an acceleration voltage of 300 KV. The samples were directly scratched from the $TiO_2$ NT layers and drop-casted onto TEM copper grids coated with a lacey carbon film.

X-ray photoelectron spectrometer (XPS, PHI 5600 XPS spectrometer, US) was used for chemical compositional analysis of the samples. XPS spectra were acquired using monochromatic X-rays with a pass energy of 23.5eV. All the XPS element peaks are shifted to the $Ti_{2p}$ standard position. Spectra acquired at different depth of $TiO_2$ NT layer were obtained by $Ar^+$ ion sputtering removal of the tube layers. The $TiO_2$ NTs were exposed to argon ion beam sputtering (3.5 kV, current density 15 mA) and the sputtered sample area was 3×3 mm. The incident angle of Argon beam on the surface of $TiO_2$ NTs is $36^o$ – this corresponds to a nominal sputter rate of ≈ 2 nm/min on a $SiO_2$ reference layer.

TOF-SIMS: Time-of-Flight Secondary Ion Mass Spectrometry (TOF-SIMS) was performed on a TOF-SIMS 5 spectrometer (ION-TOF; Münster, Germany) using 25 keV $Bi^+$ bunched down to < 0.8 ns. The primary ion dose density was $5\times10^{11}$ ions/$cm^2$ with an area of 100 µm × 100 µm using negative polarity. Spectra were normalized to their total intensity and calibrated on $CH_2^-$, $C_2^-$, $CN^-$ and $CNO^-$. Depth profiles were carried out with crater size of 250 µm × 250 µm and a measured spot of 50 µm × 50 µm in center of crater. The sputter beam used was Cs 500 eV and the measurement beam is $Bi^+$ using negative polarity.

EPR spectra were recorded on a JEOL continuous wave spectrometer JES-FA200 equipped with an X-band Gunn oscillator bridge, a cylindrical mode cavity, and a helium cryostat. The samples were measured in the solid state in an air atmosphere in quartz EPR tubes with a similar loading of ≈ 0.5 mg. Background spectra were obtained at the same measurement conditions. The spectra shown were measured on the following parameters: Temperature 7 K (liquid Helium), microwave frequency ν 8.950 GHz (1) and 8.955 GHz (2), modulation width 0.5 mT, microwave power 1 mW, modulation frequency 100 kHz, and time constant 0.1 s. Spectral simulation was performed using the program W95EPR written by F. Neese [S2].

Raman measurements were performed on a Spex 1403 Raman Spectrometer with a line of a HeNe laser (632 nm) as the excitation source.

**Photocatalytic $H_2$ evolution:**

Photocatalytic hydrogen generation was measured under open circuit conditions under AM 1.5 (100 mW/$cm^2$) solar simulator illumination. The amount of accumulated $H_2$ produced in the head space of sealed quartz tubes was measured using a Varian gas chromatograph (Shimadzu) with a TCD detector. For rate determination, data were taken in regular intervals. For measurements the $TiO_2$ nanotube layers (≈ 1 $cm^2$ active area) were put into 10 mL DI water/methanol (50/50 vol. %) solution and illuminated for 24h.

**Electrochemical and photoelectrochemical characterization:**

Impedance measurements were performed on a Zahner IM6 (Zahner Elektrik, Kronach, Germany) work station under dark conditions in 0.1M $Na_2SO_4$. The electrochemical configuration consists of a platinum grid as a counter electrode and a Haber-Luggin capillary with Ag/AgCl (3 M KCl) electrode as a reference electrode. For Mott-Schottky measurements the potential window applied was -0.1V to 0.5V capacitance data were evaluated at 1Hz. EIS measurements were carried out at 200 mV in a frequency window from 0.01Hz to 10000Hz. Impedance data were fitted using a Randle's circuit (using a constant phase element CPE: $R_{\Omega}/R_{ct}$CRE) and the Zahner IM6 simulation software.

Photoelectrochemical characterization was carried out with an electrochemical setup that consisted of a classical three-electrode configuration that is with an Ag/AgCl reference electrode and a Pt plate as a counter electrode. The illumination part of the setup consists of a 150 W Xe arc lamp (LOT-Oriel Instruments) as the irradiation source and a Cornerstone motorized 1/8m monochromator. The monochromized light was focused to a spot of 5×5 $cm^2$ onto sample surface (through a quartz pass window in the electrochemical cell). Photocurrent spectra were acquired in 0.1 M ($Na_2SO_4$) at a potential of 500 mV (Ag/AgCl). At each wavelength a photocurrent transient was acquired and the steady state photocurrent is recorded. Photocurrent transients were recorded for 20 s using an electronic shutter system and A/D data acquisition. The samples were pressed by a copper back plate against an O-ring opening on the side of an electrochemical cell, leaving 1 $cm^2$ exposed to the electrolyte.

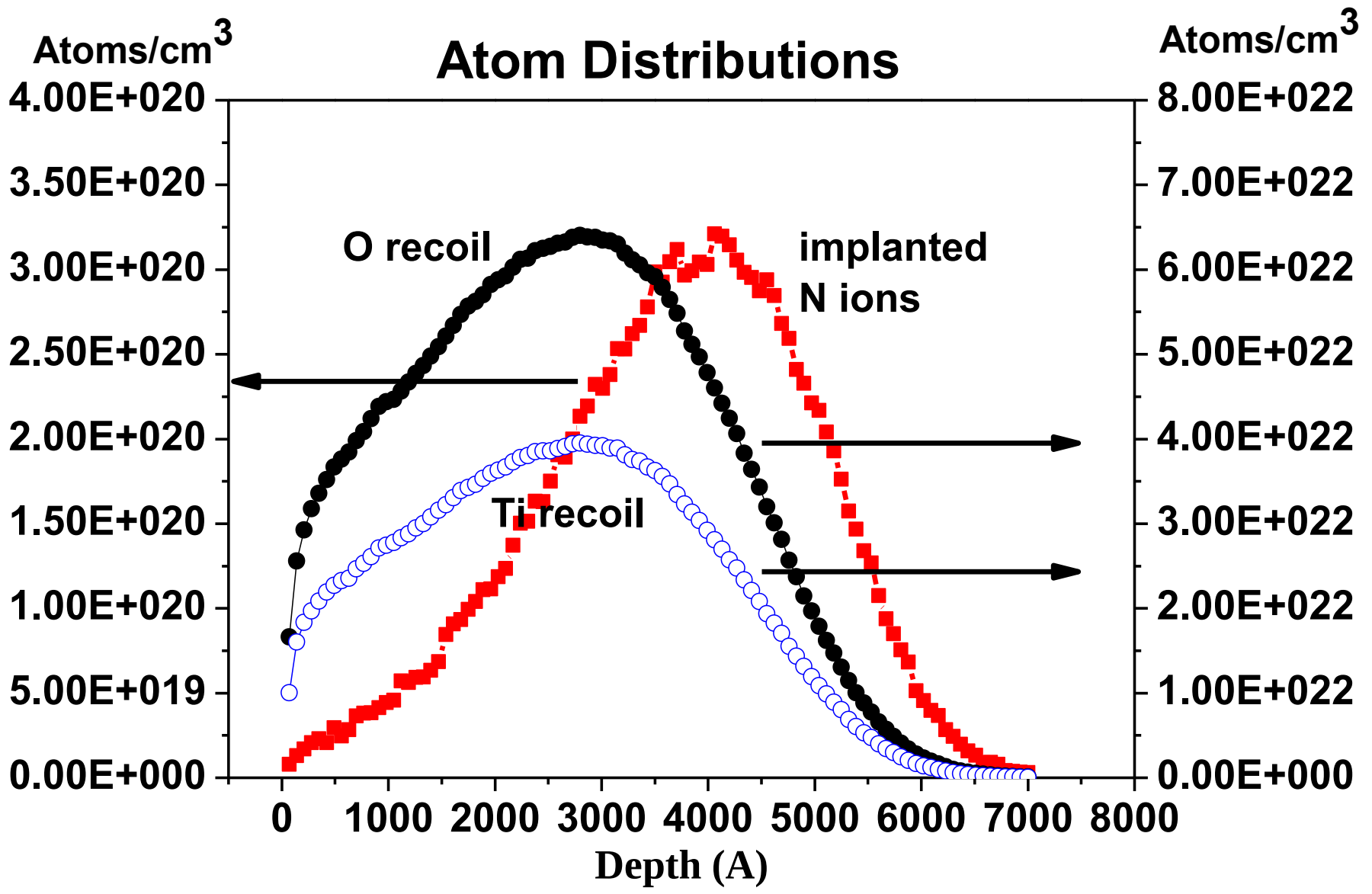

Fig.S1. Depth-distribution of N ions, Ti recoils and O recoils into a $TiO_2$ nanotube target with an energy of 60 keV at a nitrogen dose of $1\times10^{16}$ atoms/cm$^2$.

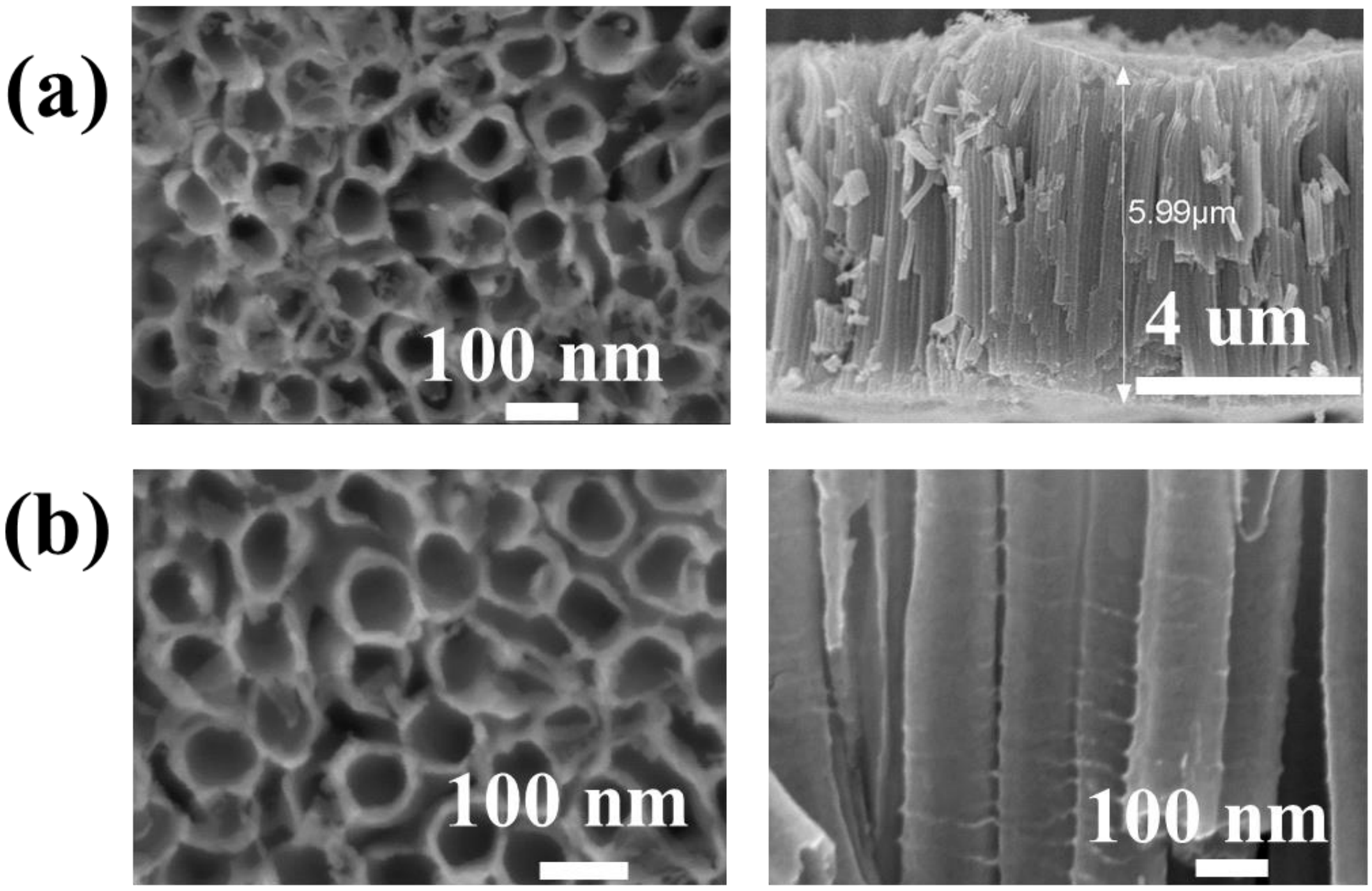

Fig. S2. SEM images for top and side view of $TiO_2$ nanotube layers (a) after low dose N implantation ($8\times10^{14}$ atoms/$cm^2$) and (b) after high dose N implantation ($1\times10^{16}$ atoms/$cm^2$).

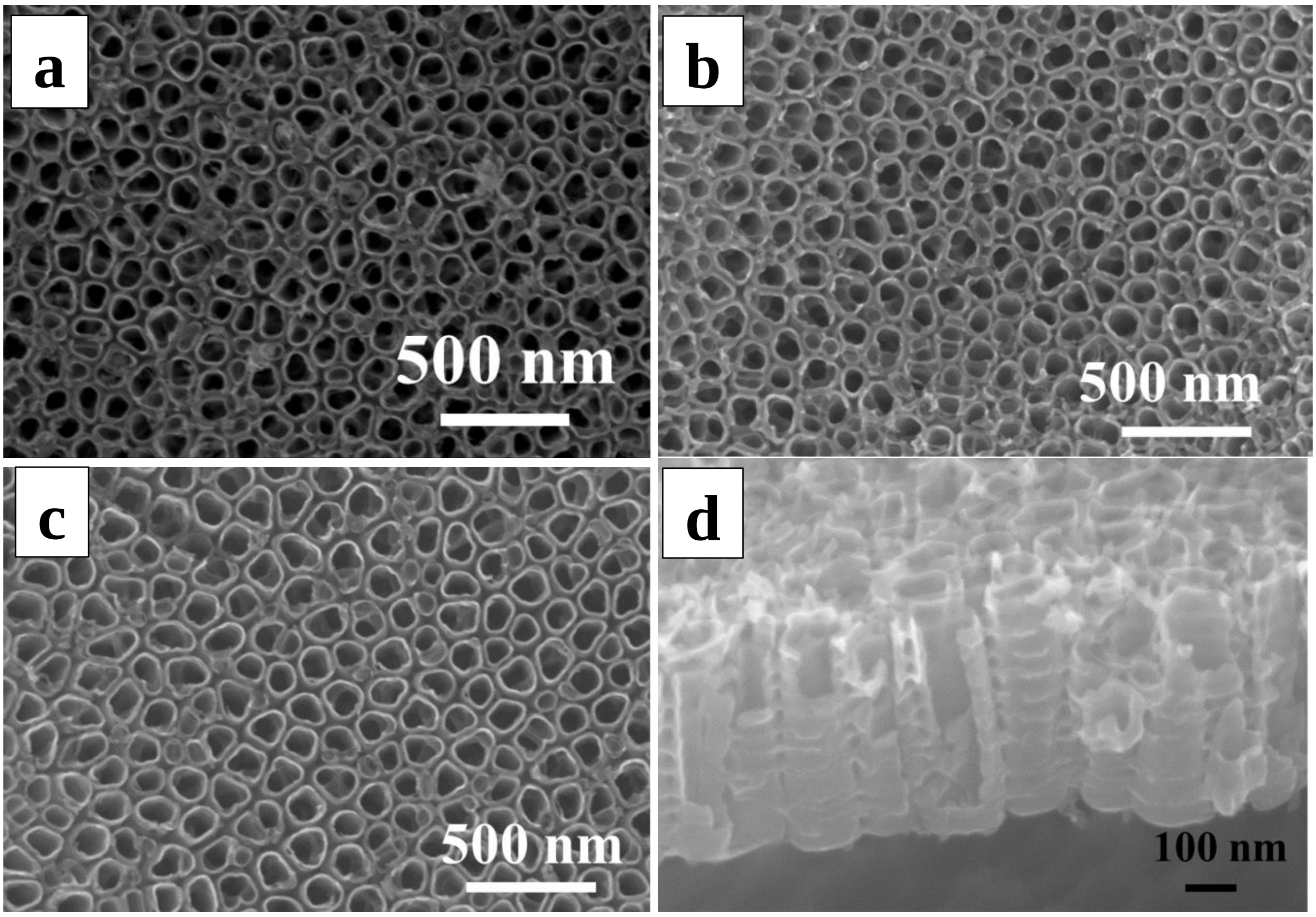

Fig. S3. SEM images for top view of $TiO_2$ short nanotube layers (a) before, (b) after low dose N implantation ($8\times10^{14}$ atoms/cm$^2$) and (c) after high dose N implantation ($1\times10^{16}$ atoms/cm$^2$). (d) Cross section view of nanotubes before implantation.

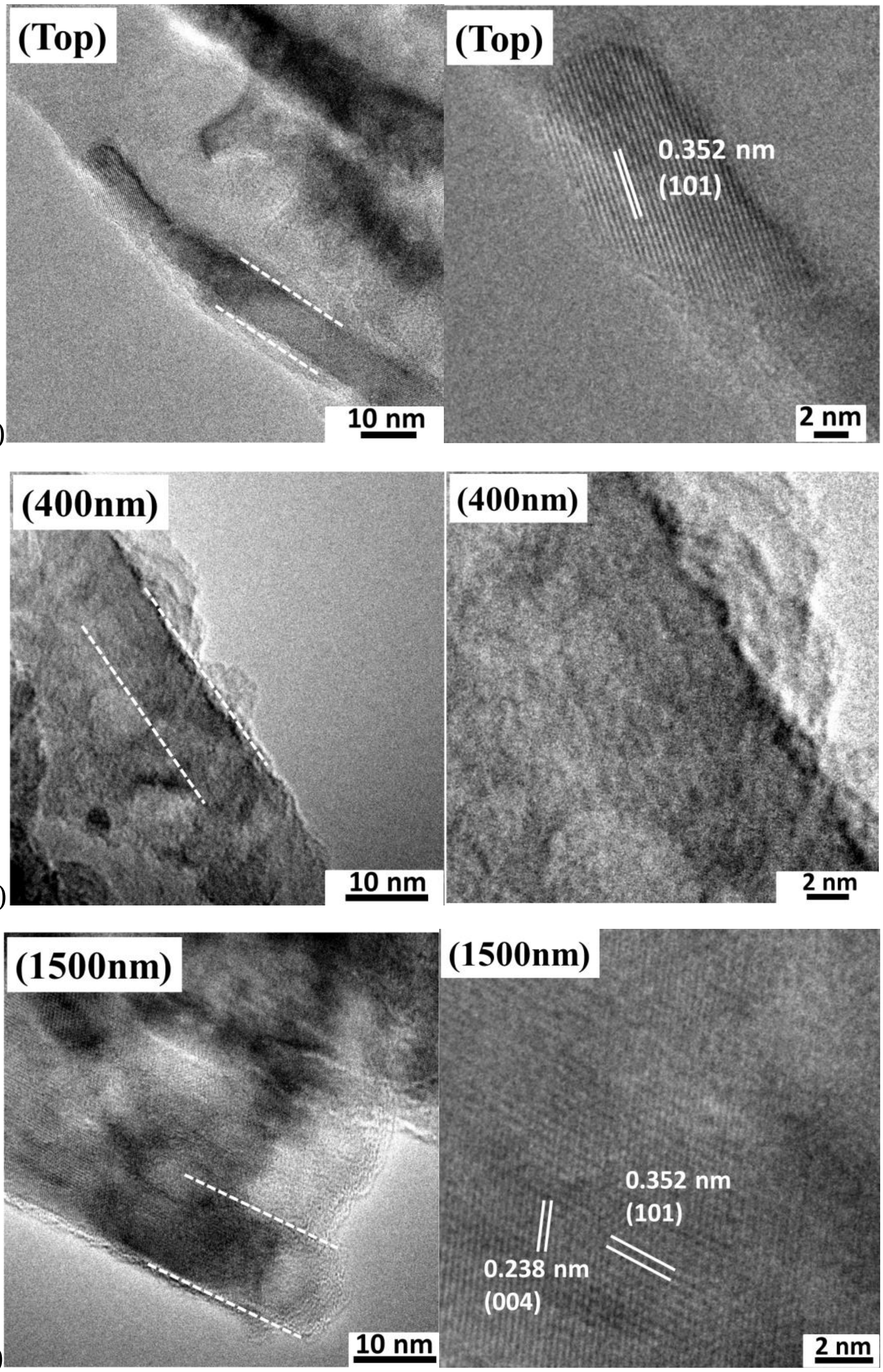

(a) (b) (c)

Fig. S4 HRTEM images of N implanted $TiO_2$ NTs with a dose of $1\times10^{16}$ ions/$cm^2$, (a) top, (b) ~ 400 nm deep and (c) ~ 1500 nm deep from top.

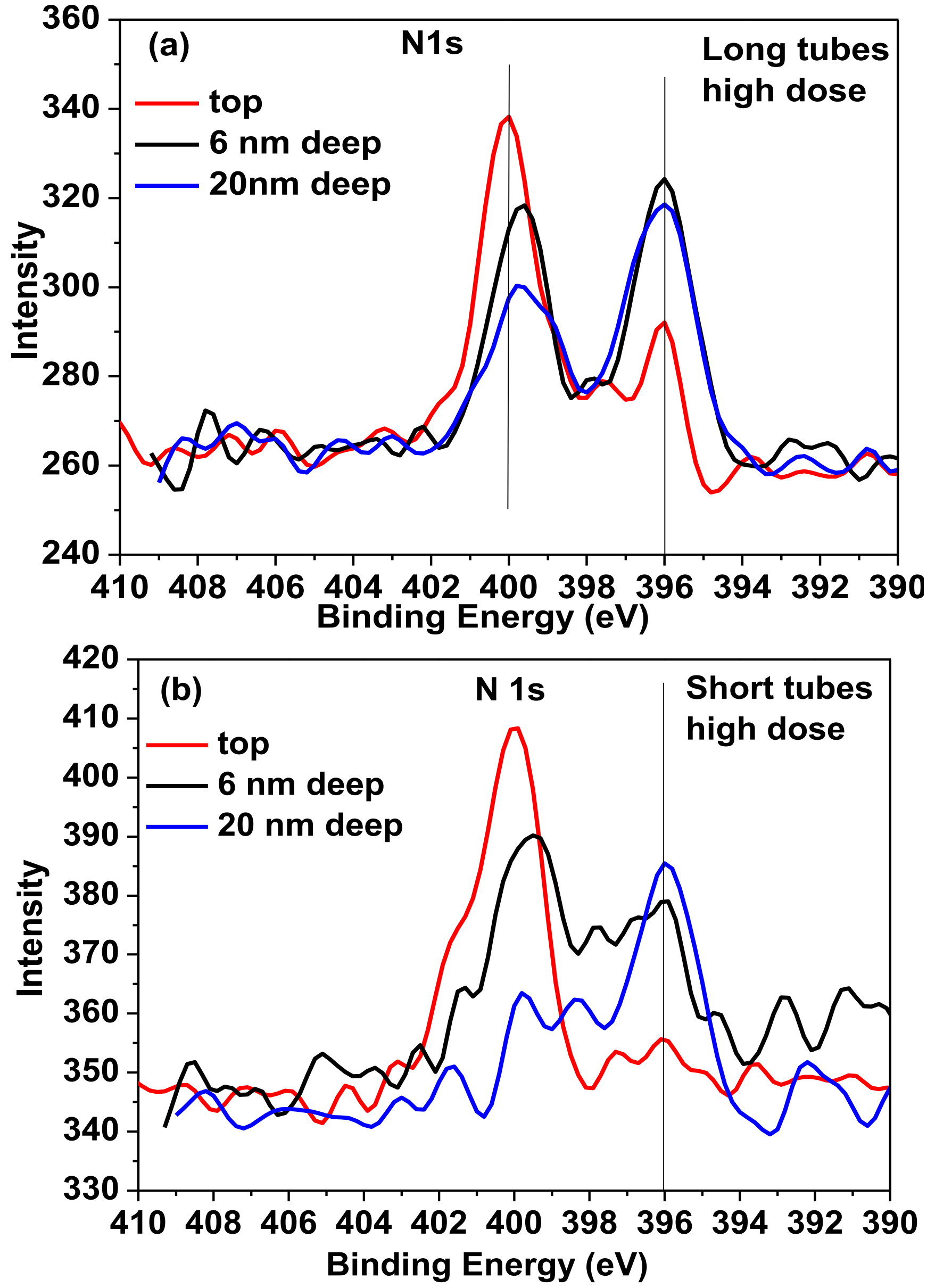

Fig. S5. XPS characterization measured at the top and different depths for N implanted long (a) and short (b) $TiO_2$ NTs with a dose of $1\times10^{16}$ atoms/cm$^2$.

If measured at top of the $TiO_2$ nanotubes, the binding energy of nitrogen is at ≈ 400 eV, considering with typically observed environmental adsorbed $N_2$. When sputter removing the top, the intensity of a clear nitrogen peak located at 396 eV becomes apparent, which suggests the implanted nitrogen in the $TiO_2$ nanotubes forms a Ti-N bond. This is observed for short and long $TiO_2$ nanotubes.

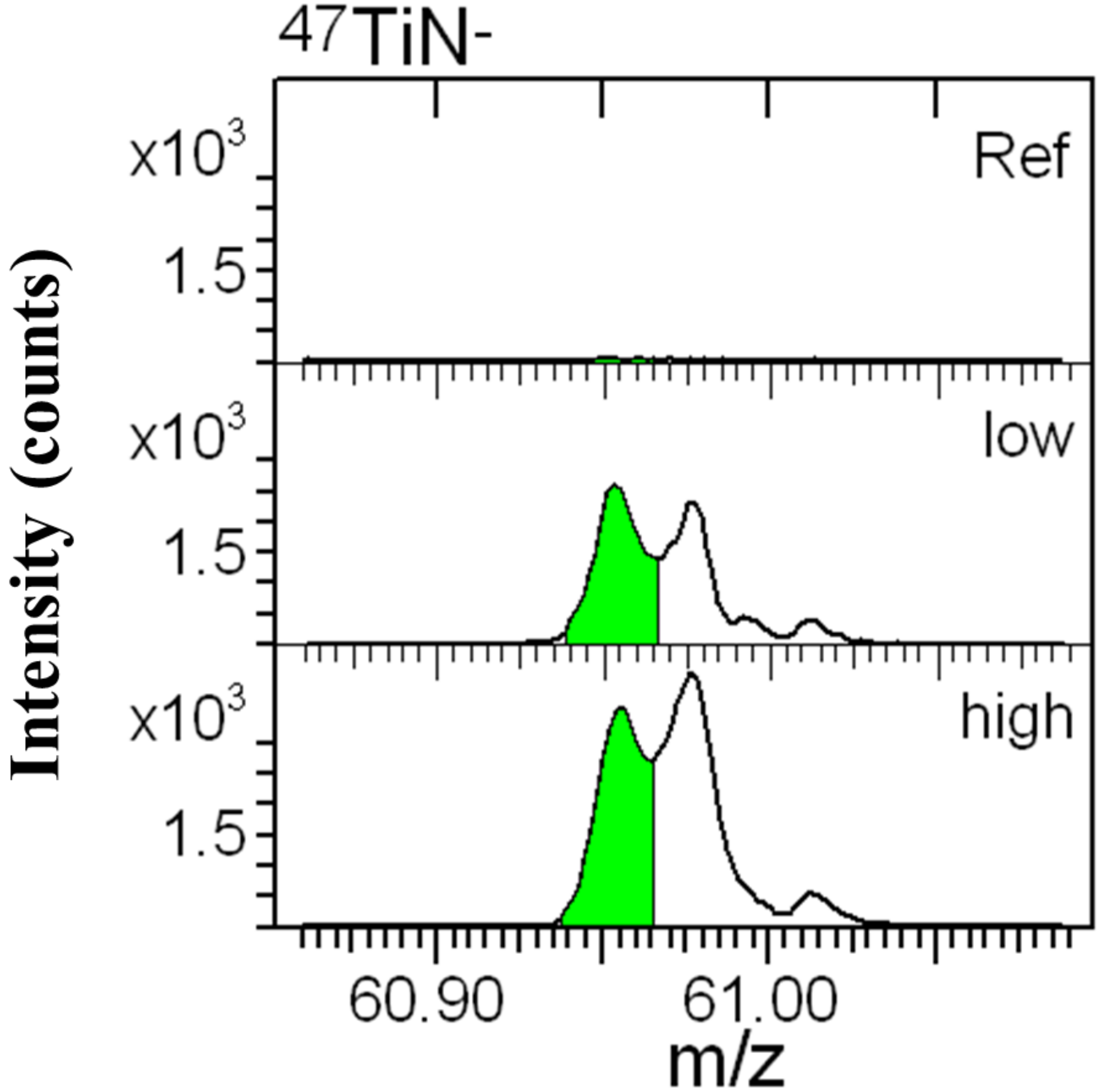

Fig. S6. TOF-SIMS spectra taken for the differently doped $TiO_2$ nanotubes.

The presence of nitrogen for lower concentrations can be confirmed by TOF-SIMS. In the low dose implanted $TiO_2$ nanotubes, a clear $^{47}$TiN- fragment becomes apparent which also appears in the high dose implanted tubes. The isotopic signal was chosen in order to avoid overlap with TiO fragments, specifically $^{46}$TiO. This signal cannot be detected in the bare $TiO_2$ nanotubes (not implanted).

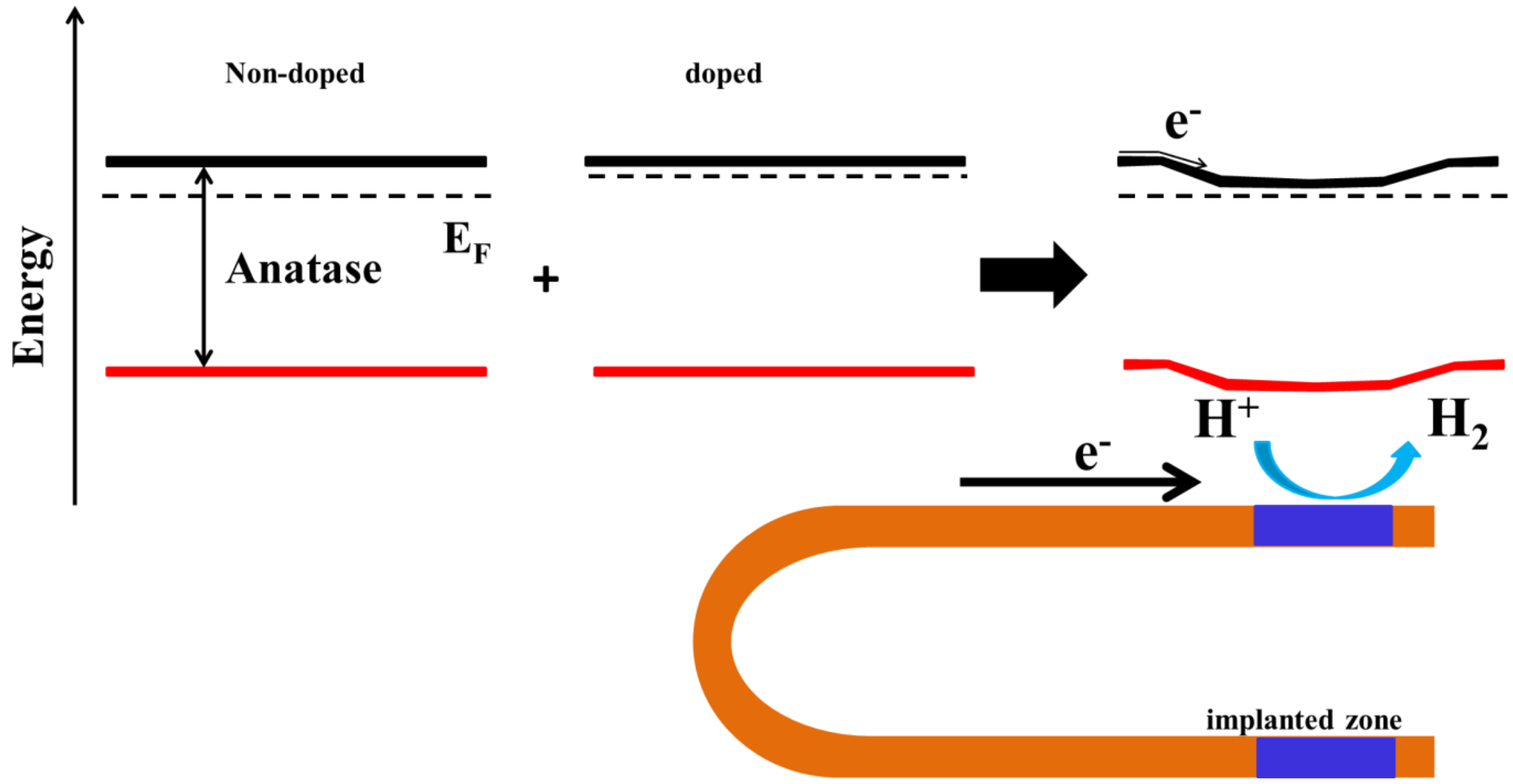

Fig. S7. Formation of a homo-junction between low dose implanted and non-implanted parts of the tubes.